\documentclass[twocolumn]{elsart}
\newcommand{\aff}[2]{Dipartimento di Fisica dell'Universit\`a #1 e 
Sezione INFN, #2, Italy.}
\usepackage[dvips]{graphicx}
\newcommand{\be}{\begin{equation}}
\newcommand{\ee}{\end{equation}}
\newcommand{\bea}{\begin{eqnarray}}
\newcommand{\eea}{\end{eqnarray}}
\newcommand{\bc}{\begin{center}}
\newcommand{\ec}{\end{center}}
\newcommand{\bt}{\begin{tabular}}
\newcommand{\et}{\end{tabular}}
\newcommand{\bfig}{\begin{figure}}
\newcommand{\efig}{\end{figure}}
\newcommand{\ei}{\end{itemize}}
\newcommand{\bleft}{\begin{flushleft}}
\newcommand{\eleft}{\end{flushleft}}
\newcommand{\bright}{\begin{flushright}}
\newcommand{\eright}{\end{flushright}}
\newcommand{\bpage}{\begin{minipage}}
\newcommand{\epage}{\end{minipage}}

\newcommand{\pip}{\ensuremath{\pi^+\,}}
\newcommand{\pim}{\ensuremath{\pi^-\,}}
\newcommand{\piz}{\ensuremath{\pi^0\,}}

\newcommand{\Eta}{\ensuremath{\eta\,}}
\newcommand{\etap}{\ensuremath{\eta'\,}}

\newcommand{\fietag}{\ensuremath{\phi\rightarrow\eta\gamma\;}}
\newcommand{\fietapg}{\ensuremath{\phi\rightarrow\eta'\gamma\;}}

\newcommand{\etapippimpiz}{\ensuremath{\eta\rightarrow\pip\pim\piz\;}}
\newcommand{\etapizpizpiz}{\ensuremath{\eta\rightarrow\piz\piz\piz\;}}

\newcommand{\etappippimeta}{\ensuremath{\etap\rightarrow\pip\pim\eta\;}}
\newcommand{\etappizpizeta}{\ensuremath{\etap\rightarrow\piz\piz\eta\;}}

\begin{document}
\begin{frontmatter}
\title{Light meson spectroscopy with the KLOE experiment.}
\author{The KLOE collaboration}\thanks{\scriptsize A.Aloisio, F.Ambrosino, A.Antonelli, M.Antonelli, C.Bacci, M.Barva, G.Bencivenni, S.Bertolucci, C.Bini, C.Bloise
% ...
Universit\`a
  ``Roma Tre'', Via della Vasca Navale, 84,
 I-00146, Roma, Italy, e-mail dimicco@fis.uniroma3.it},
\address[Roma3]{\aff{``Roma Tre''}{Roma}}

\begin{abstract}
 In this paper we describe the status of the analyses in progress
on light meson spectroscopy  in the KLOE experiment. We present the analyses of $\phi$ decays into $f_0(980)\gamma$ 
and $a_0(980)\gamma$, the Dalitz plot analysis of the $\eta \to \pi^+ \pi^- \pi^0$ decay,  the 
branching ratio measurement of $\eta \to \pi^0 \gamma \gamma$, the upper limits on $Br(\eta \to 3 \gamma)$ and $Br(\eta \to \pi^+ \pi^-)$, the measurement of the ratio $Br(
\phi \to \eta' \gamma)/Br(\phi \to \eta \gamma)$  and $\phi$ leptonic width measurements.
\end{abstract}
\end{frontmatter}
\section{Introduction}
 The KLOE detector \cite{K-dc}, operates at the Frascati $e^+e^-$ collider 
DA$\Phi$NE\ \cite{Dafne}, which runs at a CM energy $W$ equal to the $\phi$-meson 
mass, $W$$\sim$1019.5 MeV. The analyses presented here are based on data collected in the years 
2001 and 2002 for an integrated luminosity of $\sim 450~ pb^{-1}$ corresponding to 1.5 billions of $\phi$ and 
20 millions of $\eta$ mesons
[Br(\fietag) $\sim 1.3$\% \cite{PDG2002}]. This
means that  KLOE can also study  $\eta$ physics in a clean environment with high statistic.
\section{Search for $\phi\rightarrow$f$_0\gamma$ in $\pi^+\pi^-\gamma$
events.}

\begin{figure}[!htb]
\includegraphics[width=0.5\textwidth]{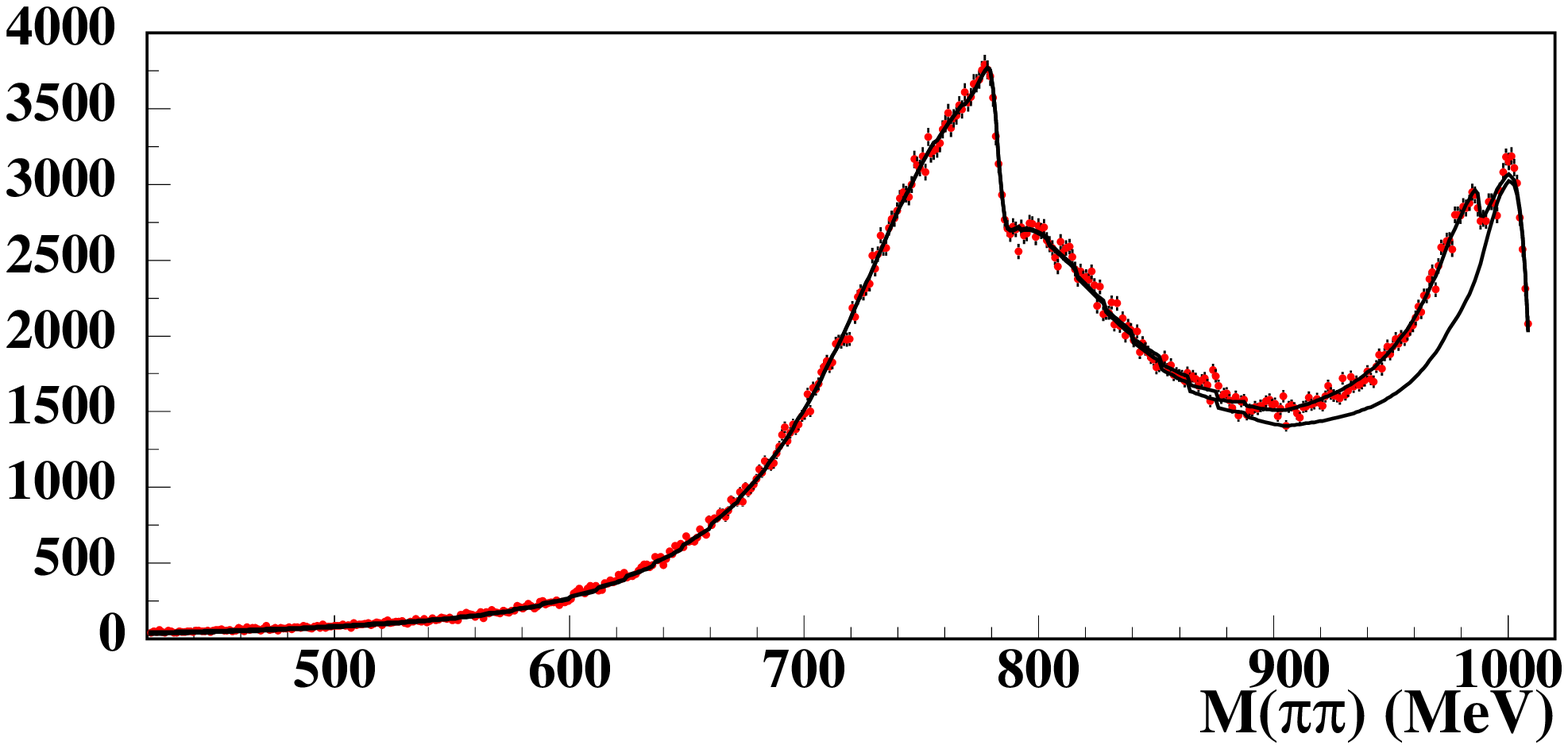}
\begin{center}
\includegraphics[width=0.5\textwidth,height=0.32\textwidth]{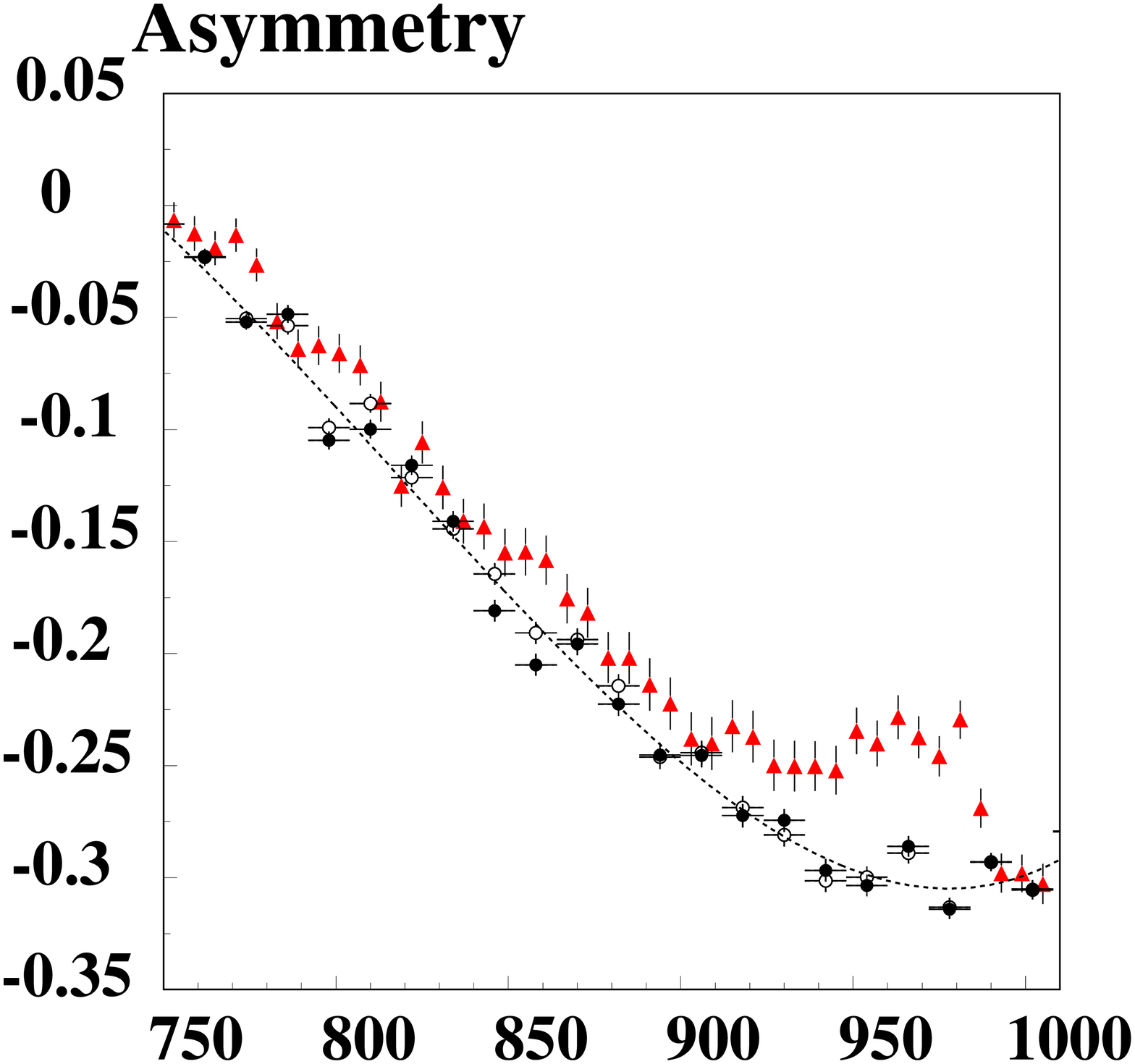}
\end{center}
\caption{ (top) $M_{\pi\pi}$ spectrum of $\pi^+\pi^-\gamma$. 
The upper (lower) curves are the result of the 
    fit and the estimated background due to ISR, FSR
    and $\rho\pi$. (bottom) Forward-Backward asymmetry $A$ as a function of $M_{\pi\pi}$. 
The curve and the black points are the
 Montecarlo expectations based on the interference between FSR and ISR
 only. The experimental data are reported as triangles.}
\label{pppm1}

\begin{flushright}
\vskip -5.2 cm 
 ${\tiny M(\pi \pi) (MeV)}$
\vskip +3.6 cm
\end{flushright}

\end{figure}

The $\phi$ radiative decays to scalar mesons, $\phi\rightarrow$S$\gamma$,
give significant insight in the assessment of the
nature of lower mass scalar mesons \cite{molti}.
With the KLOE
experiment the decays $\phi \to f_0(980)\gamma$ and $\phi \to a_0(980)\gamma$ are searched for in
$\pi^0\pi^0\gamma$ and $\eta\pi^0\gamma$ \cite{a0,f0} final states respectively.
 Moreover the $f_0(980)$ signal is also searched for $\pi^+\pi^-\gamma$ events
with a photon at large angle. The search for this signal is characterized
by the presence of a huge irreducible background due to the
initial state radiation (ISR), to 
$e^+e^-\rightarrow\pi^+\pi^-\gamma$ 
(FSR) and 
$\phi\rightarrow\rho^{\pm}(\rightarrow\pi^{\pm}\gamma)\pi^{\pm}$.
 The f$_0$ events are
searched for in the large photon angle region 45$^o<\theta<$135$^o$ to reduce ISR background. 
The f$_0$ signal appears
as a bump in the $\pi^+\pi^-$ invariant
mass M$_{\pi\pi}$ spectrum around 980 MeV. Fig.\ref{pppm1} (top) shows the spectrum obtained at $\sqrt{s}=M_{\phi}$.
\par\noindent
An overall fit to the spectrum has been done by applying the following formula:
\begin{eqnarray*}
{{dN}\over{dM_{\pi\pi}}}=&&\left[ (
{{d\sigma}\over{dM_{\pi\pi}}})_{ISR} 
+({{d\sigma}\over{dM_{\pi\pi}}})_{FSR+f_0}
+\right.\\
&& + \left. ({{d\sigma}\over{dM_{\pi\pi}}})_{\rho\pi} 
\right]\times L \times \epsilon(M_{\pi\pi})
\end{eqnarray*}
with $L$ the integrated luminosity and $\epsilon(M_{\pi\pi})$ the
selection efficiency as a function of $M_{\pi\pi}$. 
The f$_0$ amplitude is taken from the kaon-loop approach \cite{Achasov2}. 
A forward-backward asymmetry 
$A={{N^+(\theta>90^o)-N^+(\theta<90^o)}
\over{N^+(\theta>90^o)+N^+(\theta<90^o)}}$
is expected, due to the interference between FSR and ISR\cite{Binner}. Fig.\ref{pppm1} (bottom)
shows the asymmetry as a function of M$_{\pi\pi}$ 
compared to the prediction based on the ISR-FSR interference alone. 
A significant deviation from the prediction is observed in the high mass
region clearly due to the f$_0$ contribution.

\section{Dynamics of \etapippimpiz}

The dynamics of the \etapippimpiz decay has been studied with a Dalitz plot analysis. 
The conventional variables X and Y are defined as:
 $X = \sqrt{3}\frac{T_{+}-T_{-}}{Q_{\eta}}$,$Y = \frac{3 T_{0}}{Q_{\eta}}-1$, where $
Q_{\eta} = m_{\eta}-2 m_{\pip}-m_{\piz}
$ and  $T_{+}$, $T_{-}$ and $T_{0}$ are the kinetic energies of the particles. The measured distribution has been fitted as:
$
\vert A(X,Y)\vert^{2} \simeq (1 + aY + bY^{2} + cX + dX^{2}+ eXY + ...)$.
C-parity conservation prevents odd powers in $X$ being present in the
expansion: thus parameters c and e should be zero as confirmed by our fit. 
The results of the fit are shown in table \ref{tab:unica}
\begin{table}[ht]
\begin{center}
{\scriptsize
\caption{  Fitted parameters P($\chi^2$) = 52 \% of $\eta \to 3 \pi$ Dalitz plot.} 
\vskip 0.1 in
\begin{tabular}{|c|c|c|}\hline

 a & b & c \\\hline

 $-1.075\pm .008$ & $.118\pm.009$ & $-.0005\pm.004$ \\\hline
 d & e& f \\\hline
$.049\pm.008 $ & $-.004\pm .01 $ & $.13 \pm .02$ \\\hline 
\end{tabular} }\label{tab:unica}
\end{center}
\end{table}
Efficiency is $\sim 36$ \% over the whole Dalitz plot.
The evaluation of systematic effects is under completion.

\section{Rare and forbidden $\eta$ decays ($\eta \to \pi^0\gamma \gamma$,$\eta \to \pi^+ \pi^-$,$\eta \to \gamma \gamma \gamma$)}
\begin{figure}[!htb]
\begin{center}
\includegraphics[width=0.4\textwidth]{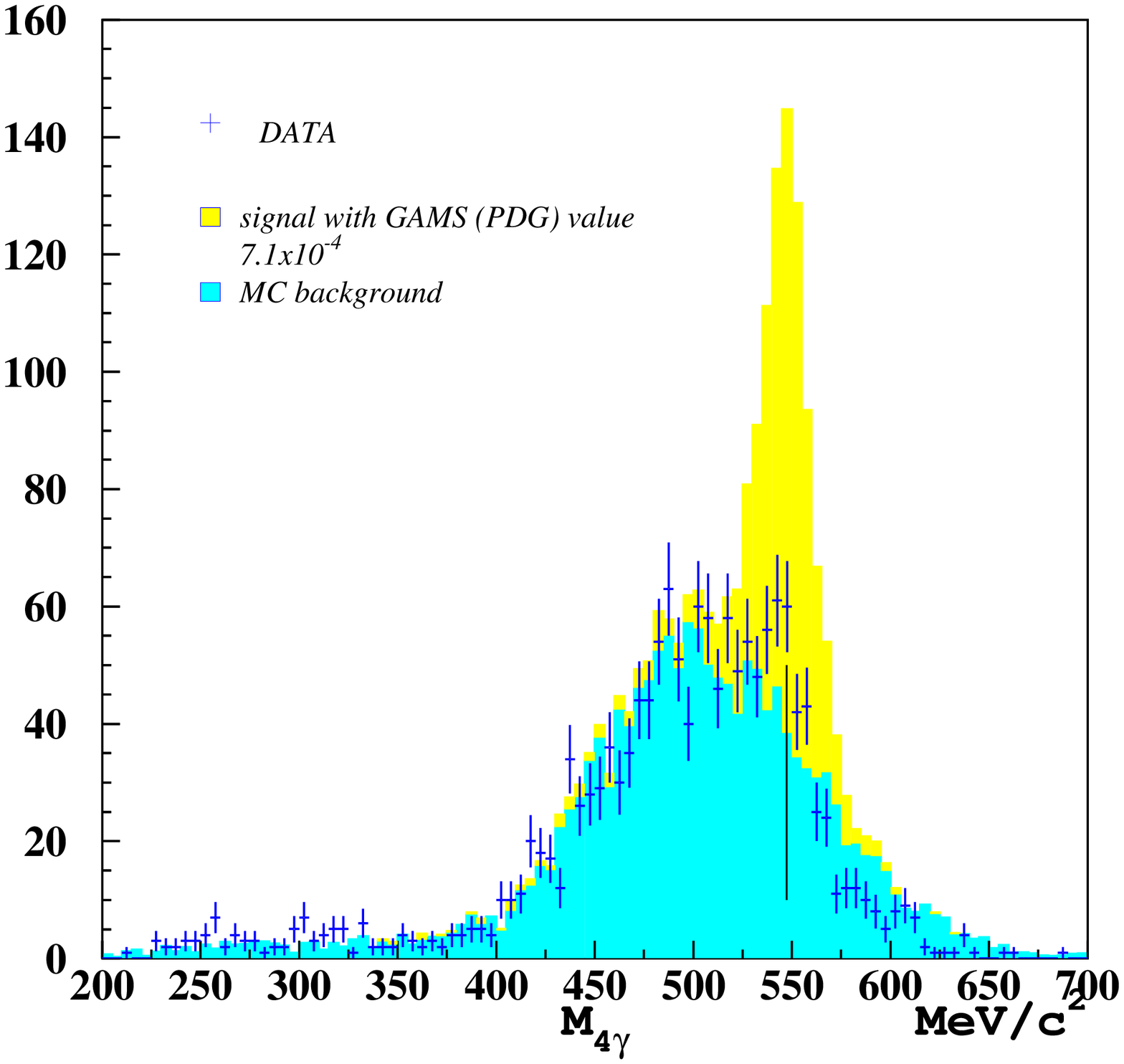}
\includegraphics[width=0.4\textwidth]{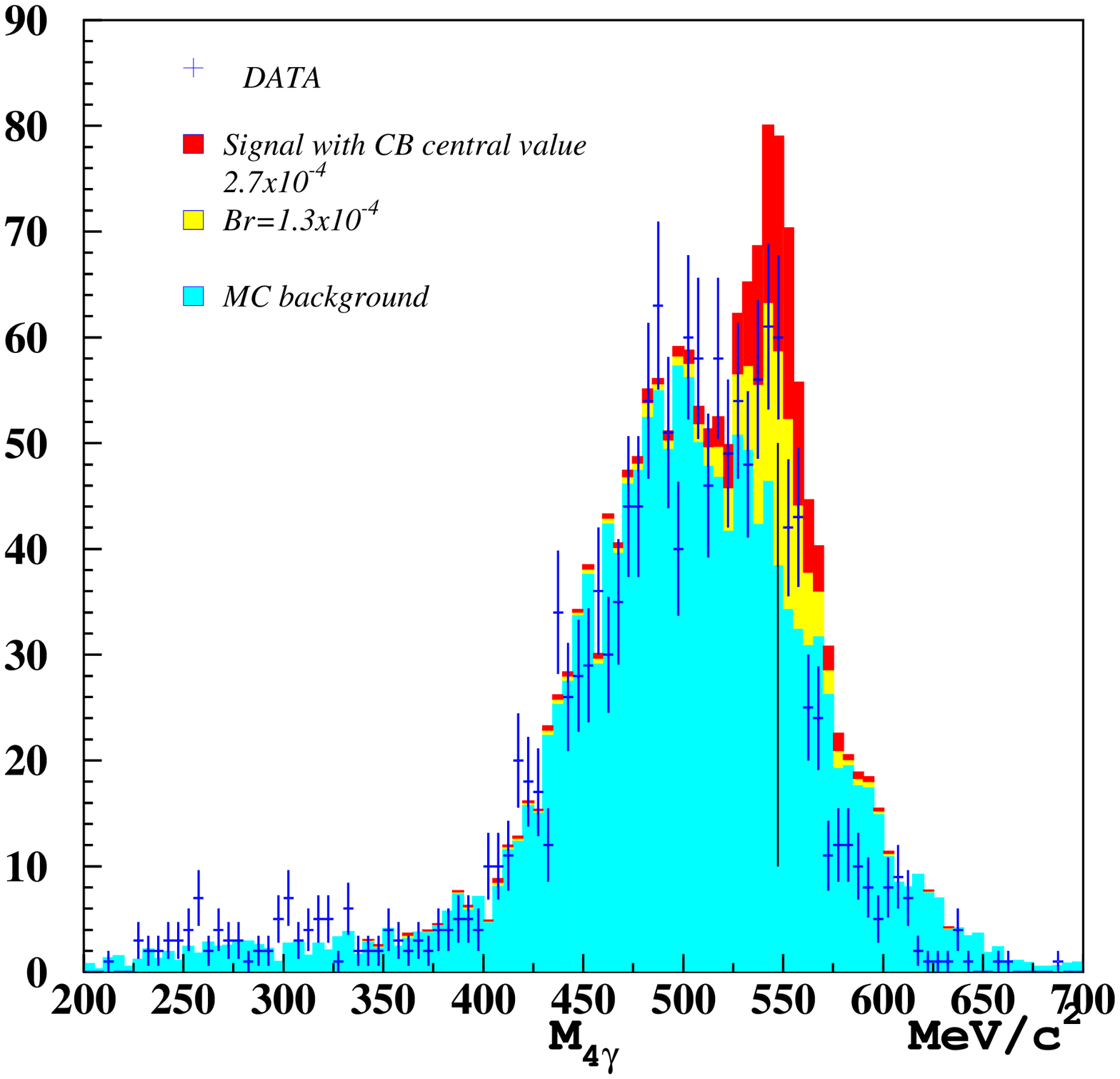}
\end{center}
\caption{ M($4\gamma$), the spectra expected from the GAMS\cite{Gams} and Crystall
Ball\cite{Crystall} measurement are shown. In the lower plot we show also the expected spectrum 
for a $Br \sim 1/2$ of C.B. result.} \label{fig:comp}
\end{figure}

The $\eta \to \pi^0 \gamma \gamma$ decay is interesting to test the Chiral Perturbation Theory prediction for the branching ratio and $m_{\gamma \gamma}$ spectrum\cite{BandG,Oset}. The most accurate measurement for the branching ratio\cite{Gams} is, infact, far from any theoretical prediction 
for this decay based on ChPT. Recently a new measurement has been performed \cite{Crystall} giving a much lower value than 
the previous one, with a larger error.  
All previuos experiments were done at hadron machines, using mainly $\pi^- p \to \eta n$, and are largely dominated by $\pi^0 \pi^0$ background and 
geometrical acceptance. KLOE performs a measurement of competitive precision in a cleaner environment. Furthermore, it has different background topologies and experimental systematics.
The signal is searched for by looking for a $\pi^0 \gamma \gamma \gamma$ topology, where the further $\gamma$ comes from $\phi \to \eta \gamma$. Five prompt clusters are required and an overall kinematic fit requiring $\pi^0$ mass is performed.
The  clusters energy must be $>$ 30 MeV and azimutal angle $>$ $21^{\circ}$ to reject fake clusters coming from 
machine background. The dominant background channel is $\eta \to 3 \pi^0$ that had been reduced with several topological cut.  
With this selection we obtain an efficiency of $~ 5.7 \%$. 
To give an idea of the sensitivity,
In fig. \ref{fig:comp} we
compare the $M(4\gamma)$ data spectrum with MC based predictions of
signal and background in two hypothesys for the size of signal: one
based on PDG value\cite{Gams} and one based on the recent CB result\cite{Crystall}.
 It is
evident that our data are incompatible with \cite{Gams} and are 
marginally in agreement with \cite{Crystall}. The background simulation and the efficiency for the signal is still under study. \\

\par\noindent $\eta \to 3 \gamma$ decay is  \emph{C} violating. It is a sensitive test of \emph{C} violation
in the strong and electromagnetic interactions. For the details of this analysis see ref.\cite{eta3g}. 
The KLOE result for the branching ratio is: $Br(\eta \to \gamma \gamma \gamma) \le 1.6 \times 10^{-5}$ @90 \% C.L. This limit is the  best experimental limit for this decay. The expected branching ratio from the Standard Model is $\le 10^{-12}$ \cite{Herczeg},
so any discovery of a larger decay rate would be a clear signal of Standard Model deviation. \\

\par\noindent $\eta \rightarrow \pi^+\pi^-$ decay is  P and CP violating. 
This
decay is allowed as a weak direct CP violating 
decay  with a very low branching ratio: BR($\eta \rightarrow \pi^+\pi^-$)$\sim
10^{-27}$ \cite{Shab}. Therefore the detection of this decay at an accessible level would be a signal of P
and CP violation not explainable in the Standard
Model framework. The latest published \cite{russilimite} direct search of this decay has given 
the following 90\% C.L. upper limit: $BR(\eta \rightarrow \pi^+\pi^-)< 3.3\times 10^{-4}$.
In KLOE the signal is searched for the M($\eta$) region of the  
$\pi^+\pi^-$ invariant mass spectrum of $\pi^+\pi^-\gamma$ events
selected according to the f$_0$(980)$\rightarrow\pi^+\pi^-$ analysis
described before (see fig. \ref{pppm1}). 
The signal efficiency is: $\epsilon_s$=16.6\%. 
The expected signal has a Gaussian shape with a
mass resolution of 1.33 MeV.
No signal is observed. The background is determined by fitting the theoretical model 
for $\pi^+ \pi^- \gamma$ sample to the full spectrum. In order to determine an upper limit, we have added to this background 
a Gaussian function representing the signal multiplied by a constant $N_s$. 
We obtain: $N_s=-8\pm 24$.
The 90\% confidence level upper limit on the number of events
is obtained using the tables in \cite{FeldmannCousins}: $N_s < 33$. The 
branching ratio is $BR(\eta\rightarrow\pi^+\pi^-)={{N_s}\over{\epsilon_s N_{\eta}}}$ 
with $N_{\eta}$ the number of $\eta$ in the sample $(1.55\times 10^7)$. The
90\% C.L. upper limit is: $BR(\eta\rightarrow\pi^+\pi^-) < 1.3\times 10^{-5}$.
It improves by a factor $\sim$ 30 the current PDG limit.
\section{\Eta - \etap mixing}
Here we present the $ R = \frac{\Gamma(\fietapg)}{\Gamma(\fietag)}$ measurement.
The \etap is identified via
the decays: \fietapg ; \etappippimeta ; \etapizpizpiz and the decays
\fietapg ; \etappizpizeta ; \etapippimpiz. The final state is thus
charachterized by two charged pions and seven photons, and has no physics
background with the same topology in KLOE. After background subtraction
 we observe $3405 \pm 61 \pm 31$  \fietapg
events. We normalize to the
number of observed \etapizpizpiz decays in the same runs to obtain a
preliminary measurement of the ratio of BR's: $R = (4.9\pm 0.1_{stat} \pm 0.2_{syst})\times 10^{-3}$.
This result compares favourably with our previous estimate \cite{PLBetap}
(which already dominates
the world average \cite{PDG2002}) but with considerably improved accuracy.
\section{A new measurement of the $\phi$ leptonic width.}

KLOE has performed a new measurement of the $\phi$ leptonic
widths $\Gamma_{ll}$ with $l=e,\mu$ \cite{Dreucci}, using the two data samples taken below
($\sqrt{s}$=1017 MeV) and above ($\sqrt{s}$=1022 MeV) the $\phi$
peak together with the data taken at the $\phi$ peak. 
The dependences on $\sqrt{s}$ of the forward-backward asymmetry of Bhabha
events $A_{FB}$ and of
the $e^+e^-\rightarrow\mu^+\mu^-$ cross-section $\sigma(\mu\mu)$ around
the $\phi$ peak are sensitive to the
value of $\Gamma_{ee}$ and 
$\sqrt{\Gamma_{ee}\Gamma_{\mu\mu}}$
respectively.
We measure the $\phi$ mass $M_{\phi}$, the forward-backward asymmetry at
$W=M_{\phi}$ $A_{FB}^0$, and finally  $\Gamma_{ee}$.
The result for $\Gamma_{ee}$ is:
$\Gamma_{ee}=1.32\pm0.05_{stat}\pm0.03_{syst}~{\rm keV}$
 The result for $\sqrt{\Gamma_{ee}\Gamma_{\mu\mu}}$ is : $\sqrt{\Gamma_{ee}\Gamma_{\mu\mu}}=1.320\pm0.018_{stat}\pm0.017_{syst}~{\rm 
keV}$.
 The two results are in good
agreement consistently with lepton universality. Combining them we get:
$\Gamma_{ll}=1.320\pm0.017_{stat}\pm0.015_{syst}~keV$
with a total uncertainty below 2 \%. We point out that the value of
$\Gamma_{ee}$ is necessary for $\phi$ decay branching ratio
measurements, and play also a role in the evaluation of the hadronic
contribution to vacuum polarization \cite{Eidel}. 

\end{document}